\documentclass[reprint, superscriptaddress, longbibliography, floatfix, twocolumn, nofootinbib, amsmath, amssymb,
 aps, showkeys]{revtex4-2}

\usepackage{mathtools}
\usepackage{nicefrac}
\usepackage{braket}
\usepackage{siunitx}
\usepackage{hyperref}
\usepackage{gensymb}
\usepackage{xcolor}
\usepackage{graphicx}
\usepackage{dcolumn}
\usepackage{bm}
\usepackage{dsfont}
\usepackage{float}
\usepackage[most]{tcolorbox}
\usepackage{comment}
\usepackage{overpic}
\usepackage{soul}

\begin{document}

\preprint{APS/123-QED}

\title{
An asymmetric atom-photon architecture for device-independent quantum key distribution over 25\,km}

\author{Jonas Meiers}
\affiliation{Fachrichtung Physik, Universit\"at des Saarlandes, 66123 Saarbr\"ucken, Germany}
\affiliation{Zentrum für Quantentechnologien (QuTe), Universit\"at des Saarlandes, 66123 Saarbr\"ucken, Germany}

\author{Christian Haen}
\affiliation{Fachrichtung Physik, Universit\"at des Saarlandes, 66123 Saarbr\"ucken, Germany}
\affiliation{Zentrum für Quantentechnologien (QuTe), Universit\"at des Saarlandes, 66123 Saarbr\"ucken, Germany}

\author{Max Bergerhoff}
\affiliation{Fachrichtung Physik, Universit\"at des Saarlandes, 66123 Saarbr\"ucken, Germany}
\affiliation{Zentrum für Quantentechnologien (QuTe), Universit\"at des Saarlandes, 66123 Saarbr\"ucken, Germany}

\author{Pascal Baumgart}
\affiliation{Fachrichtung Physik, Universit\"at des Saarlandes, 66123 Saarbr\"ucken, Germany}
\affiliation{Zentrum für Quantentechnologien (QuTe), Universit\"at des Saarlandes, 66123 Saarbr\"ucken, Germany}

\author{Tobias Bauer}
\affiliation{Fachrichtung Physik, Universit\"at des Saarlandes, 66123 Saarbr\"ucken, Germany}
\affiliation{Zentrum für Quantentechnologien (QuTe), Universit\"at des Saarlandes, 66123 Saarbr\"ucken, Germany}

\author{Christoph Becher}
\affiliation{Fachrichtung Physik, Universit\"at des Saarlandes, 66123 Saarbr\"ucken, Germany}
\affiliation{Zentrum für Quantentechnologien (QuTe), Universit\"at des Saarlandes, 66123 Saarbr\"ucken, Germany}

\author{J\"urgen Eschner}
\email{juergen.eschner@physik.uni-saarland.de}
\affiliation{Fachrichtung Physik, Universit\"at des Saarlandes, 66123 Saarbr\"ucken, Germany}
\affiliation{Zentrum für Quantentechnologien (QuTe), Universit\"at des Saarlandes, 66123 Saarbr\"ucken, Germany}

\date{\today}

\begin{abstract}
Device-independent quantum key distribution (DIQKD) can guarantee security without trusting the internal workings of the measurement devices, but extending it to fiber networks demands high-quality entanglement, reliable heralding, and faithful photon transmission simultaneously. Here, we address these requirements in an event-ready asymmetric atom-photon architecture implemented over $25\,\mathrm{km}$ of spooled telecom fiber. A single trapped $^{40}\mathrm{Ca}^{+}$ ion forms one party of the protocol while a single transmitted photon measured at the remote station forms the other. 
Double quantum frequency conversion and active polarization stabilization preserve the atom-photon quantum correlation across the full link. We obtain a Clauser-Horne-Shimony-Holt parameter of $|S|=2.75^{+0.16}_{-0.15}$, exceeding the threshold $|S|=2.362$ required for a positive asymptotic secret-key fraction under the DIQKD model used. Within the same asymptotic model, this corresponds to a conservative estimate of 69 secret-key bits out of 10908 detected Bell states. The demonstrated architecture establishes a route towards DIQKD in heterogeneous, repeater-compatible quantum networks.
\end{abstract}

\maketitle

Quantum networks \cite{Kimble_2008, Wehner2018, Cirac_1999} distribute quantum information between distant nodes and hold the promise of information-theoretically secure communication through quantum key distribution (QKD) \cite{GisinQKD}. In QKD, two remote parties establish a shared cryptographic key by exchanging quantum states over an otherwise insecure channel, with security guaranteed by the laws of quantum mechanics rather than by computational hardness assumptions. A leading approach to realizing such networks relies on matter-based nodes, for example trapped ions \cite{Nadlinger_2022, Kucera2024} or neutral atoms \cite{Zhang_2022}, that store and process quantum information locally and exchange it via single photons over optical fiber, making use of existing telecommunication infrastructure. Because many of these platforms provide a natural interface to photonic polarization qubits, polarization encoding is widely used to link matter and photonic degrees of freedom. Among available platforms, trapped ions are particularly well suited for networking \cite{Nigmatullin_2016, Krutyanskiy_2023_2, Krutyanskiy_2023, OReilly_2024, Bergerhoff2024, Liu_2026, Kurz2016} owing to the high degree of control demonstrated over their internal and motional quantum states. In particular, $^{40}\mathrm{Ca}^{+}$ ions emit and absorb photons at wavelengths compatible with entangled photon-pair sources \cite{Arenskoetter_2024} and with high-efficiency conversion to the telecom band to minimize fiber transmission loss \cite{Bock_2018, Arenskoetter2023, Bock2024}. While polarization encoding is susceptible to environmentally induced drifts during fiber transmission, active polarization stabilization has enabled long-distance, long-term operation with high fidelity \cite{Kucera2024}.

The security of a QKD protocol rests on the assumptions made about the devices used to prepare and measure quantum states. Early protocols such as BB84 \cite{BB84} assume that these devices behave exactly as specified, an assumption that is difficult to certify in practice and that leaves realistic implementations vulnerable to side-channel attacks \cite{sidechannel}. Entanglement-based protocols such as E91 \cite{E91} relax this requirement by certifying non-classical correlations through the violation of a Bell inequality, but conventional implementations still rely on detailed device models for the measurement apparatus. Device-independent QKD (DIQKD) protocols \cite{Acin} remove this trust requirement by certifying security directly from the observed input-output statistics under substantially reduced device assumptions, including measurement independence and the absence of unauthorized signaling between the parties.

Realizing DIQKD imposes stringent requirements on the experimental system, including high detection efficiency, high-fidelity entangled states, randomized local measurements, and strict control of information flow between the parties, ideally over links compatible with deployed telecommunication fiber \cite{Zhang_2022,Zapatero2023}. Photon loss during transmission is a central practical obstacle, as it opens the detection loophole that DIQKD protocols must close. Experimental implementations based on remote matter qubits have established Bell-certified key generation and DIQKD-compatible correlations on trapped-ion and neutral-atom platforms \cite{Nadlinger_2022, Zhang_2022}. More recent matter-based experiments have extended these concepts to kilometer-scale fiber links, reporting DIQKD analyses accounting for finite data statistics at 10\,km and 11\,km, with positive asymptotic key rates up to 101\,km and 100\,km, respectively \cite{Liu_2026, Lu_2026}. Long-distance entanglement distribution and network automation have separately been demonstrated over metropolitan fiber links \cite{bersin2023, Strobel2024, Craddock2024, stolk2024}.

The DIQKD implementations discussed above \cite{Nadlinger_2022, Zhang_2022, Liu_2026, Lu_2026} share a matter-matter architecture in which the Bell-test outcomes at both parties are obtained from local measurements on matter qubits, requiring a quantum-memory platform at each measurement station. An asymmetric matter-photon architecture mitigates this requirement by directly measuring the transmitted photonic qubit at one side of the link. This reduces the hardware requirements at the remote measurement station and allows its placement to follow the network topology rather than its co-location with another matter-qubit platform. The architecture is therefore naturally compatible with heterogeneous and repeater-based quantum networks \cite{Bergerhoff2024, Bergerhoff2026}. 

\begin{figure*}[t]
\includegraphics[width=\textwidth]{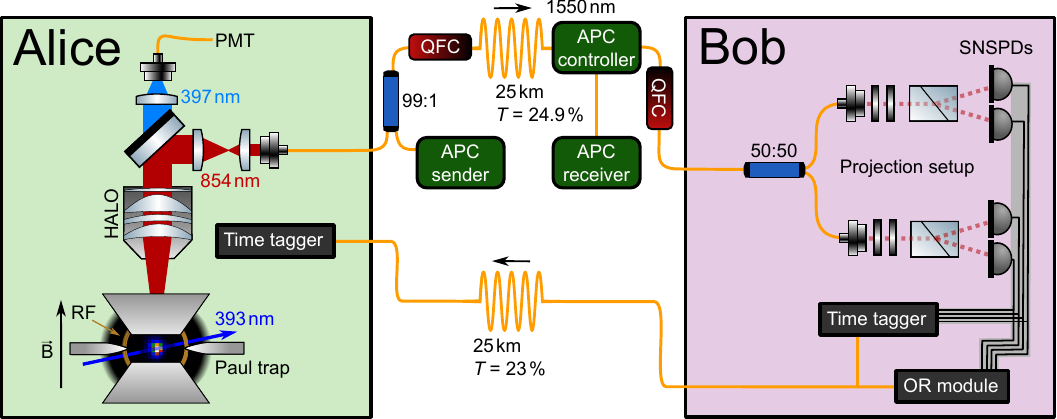}
\caption{Schematic of the atom-photon QKD link. A trapped $^{40}\mathrm{Ca}^{+}$ ion emits $854$\,nm photons that are frequency-converted to the telecom C-band ($1550$\,nm), transmitted over $25$\,km of fiber, converted back to $854$\,nm, and analyzed in polarization using single-photon detectors (SNSPDs). A separate classical fiber distributes White Rabbit timing and transmits electronic heralding signals from Bob to Alice to trigger the ion readout. Active polarization stabilization (APC) is applied on the quantum fiber to compensate slow drifts. For more details, see text. PMT: photomultiplier tube; HALO: high-numerical-aperture laser objective; RF: radio frequency; APC: automated polarization drift compensation; QFC: quantum frequency conversion; SNSPD: superconducting nanowire single-photon detector.}
\label{fig:Setup}
\end{figure*}

In this work, we implement an event-ready \cite{Rosenfeld, Sangouard_2013} atom-photon architecture for DIQKD based on entanglement between a single trapped $^{40}\mathrm{Ca}^{+}$ ion and a photon transmitted through 25\,km of optical fiber. The term event-ready refers to an electronic herald that signals the detection of a photon at the remote station and triggers a subsequent atomic readout. The transmitted photon is polarization-entangled with the ion and is frequency-converted twice, first from 854\,nm to the telecom C-band for low-loss transmission and then back to the ionic wavelength for detection. Active polarization stabilization of the fiber connection ensures stable high-fidelity entanglement. We follow the measurement protocol of \cite{Schwonnek_2021} that projects the atom-photon state randomly on a combination of one out of four atomic bases and one of two photonic bases. We use the time-resolved data to find the relevant atom-photon correlations and show that they correspond to a positive asymptotic secret-key fraction within the model of \cite{Schwonnek_2021}.

The asymmetric matter-photon architecture requires a minimal photonic detection setup which may be placed flexibly at a remote network node. This is facilitated by the integration of active polarization stabilization. 
By employing double quantum frequency conversion, our implementation nevertheless preserves compatibility with further $^{40}\mathrm{Ca}^{+}$-based network interfaces. 
The characterization of our experimental apparatus certifies it as a DIQKD link and establishes a route towards its application in heterogeneous, repeater-compatible quantum networks. 

\section{Experimental setup}\label{sec:setup}

The experimental setup, shown in Fig.~\ref{fig:Setup}, consists of two stations, Alice and Bob, connected by an optical fiber link. Alice houses the trapped-ion system, in which atom-photon entanglement is generated using a single $^{40}\mathrm{Ca}^{+}$ ion as quantum memory, while Bob performs the photonic measurement. The fiber link includes two quantum frequency conversion stages and active polarization stabilization.

Alice’s measurement setting is selected by coherently rotating the ion state with an RF pulse after photon emission and prior to readout, with the rotation angle determined in real time by a local quantum random number generator. Bob’s measurement setting is selected passively in the photonic projection stage by a $50{:}50$ beam splitter that routes each photon to one of two polarization-analysis modules of the projection setup. 

Below we present the relevant parts of the experiment; more details are provided in \cite{Bergerhoff2024}.

\begin{figure}[H]
\includegraphics[width=\linewidth]{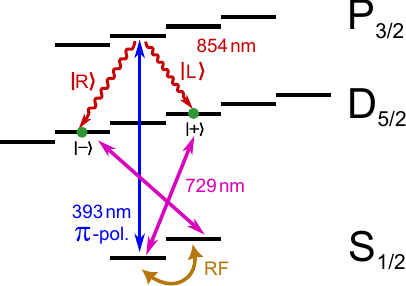}
\caption{Scheme showing relevant $^{40}\mathrm{Ca}^{+}$ levels and transitions used for atom-photon entanglement generation and atomic state readout. Excitation from $\mathrm{S}_{1/2}$ to $\mathrm{P}_{3/2}$ (blue arrow) prepares the excited state, subsequent spontaneous decay to $\mathrm{D}_{5/2}$ generates a single photon (wavy red arrows) whose polarization is entangled with the Zeeman sublevel in $\mathrm{D}_{5/2}$. Subsequent laser pulses (purple arrows) coherently transfer the atomic qubit state from $\mathrm{D}_{5/2}$ to $\mathrm{S}_{1/2}$, where an RF pulse with variable phase rotates the measurement basis. Final state readout employs shelving pulses and state-selective fluorescence (see text).}
\label{fig:scheme}
\end{figure}

\subsection{Atom-photon entanglement}\label{sec:APE_setup}

The implementation of the DIQKD protocol relies on the generation of polarization-entanglement between a single trapped $^{40}\mathrm{Ca}^{+}$ ion and an emitted photon. The ion is initially excited from the $\mathrm{S}_{1/2}$ ground state to the $\mathrm{P}_{3/2}$ excited state using a $\pi$-polarized $393$\,nm laser pulse. Spontaneous decay into the metastable $\mathrm{D}_{5/2}$ manifold leads to the emission of a single $854$\,nm photon, as illustrated in Fig.~\ref{fig:scheme}.  
Using a high-numerical-aperture laser objective (HALO, $\mathrm{NA}=0.4$) aligned along the magnetic-field quantization axis, we collect the left- and right-circularly polarized photons emitted in the decay channels to the states $\ket{+}=\ket{\mathrm{D}_{5/2},+1/2}$ and $\ket{-}=\ket{\mathrm{D}_{5/2},-3/2}$, respectively. This process generates entanglement between the internal state of the ion and the polarization of the emitted photon.

Unequal Clebsch-Gordan coefficients in the spontaneous emission process lead to an imbalance in the raw entangled state. We compensate for this imbalance by an additional preparation step that probabilistically discards population in the $\ket{-}$ state, resulting in the maximally entangled state
\begin{align}
\ket{\psi} &= \sqrt{\tfrac{1}{2}} \left( \ket{+}\ket{\mathrm{L}} + e^{i \omega_L t}\ket{-}\ket{\mathrm{R}} \right) ,
\label{eq: ape}
\end{align}
where $\ket{\mathrm{L}}$ and $\ket{\mathrm{R}}$ denote left- and right-circular photon polarizations with respect to the quantization axis. The phase factor $e^{i\omega_L t}$ arises from Larmor precession of the atomic qubit state, with Larmor frequency $\omega_L \approx 2\pi \times 9.6\,$MHz, during the time interval $t$ between photon emission and detection. This phase causes the measured atom-photon correlations to oscillate with the detection time (see Sec.~\ref{sec:results}). 

Coherent transfer of the atomic qubit between the $\mathrm{S}_{1/2}$ and $\mathrm{D}_{5/2}$ manifolds is performed with a narrow-linewidth 729-nm laser. Additional lasers at 397\,nm and 866\,nm (not shown in Fig.~\ref{fig:scheme}) are used for Doppler cooling and state-dependent fluorescence readout. Fluorescence photons at $397$\,nm are collected with the same objective as 854\,nm photons and detected using a photomultiplier tube (PMT). All detection signals are recorded by a time-tagging system.

Radio frequency (RF) pulses resonant with the $\ket{\mathrm{S}_{1/2},-1/2} \leftrightarrow \ket{\mathrm{S}_{1/2},+1/2}$ transition are used to apply controlled rotations on the ground-state Zeeman qubit. The RF phase, which selects the measurement basis for Alice in the QKD protocol, is controlled by a quantum random number generator. During the electronic latency between a photon detection at Bob and the subsequent atomic readout at Alice, a spin-echo sequence is applied to the ion qubit to reduce dephasing and preserve atomic coherence.

\subsection{Fiber link and single-photon processing}

The collected $854$\,nm photons are coupled into a single-mode fiber and transmitted to a quantum frequency conversion (QFC) stage, where their wavelength is shifted to the telecom C-band at $1550$\,nm \cite{Bock_2018,Arenskoetter2023,Bock2024}. After transmission through the 25\,km-long optical fiber link, the photons are converted back to $854$\,nm by a second QFC and routed to the photonic projection setup. This setup uses a $50{:}50$ fiber beam splitter for random basis selection. Each output is followed by a free-space polarization-analysis stage consisting of a half-wave plate, a quarter-wave plate, and a Wollaston prism. The resulting four photonic detection paths are coupled to superconducting nanowire single-photon detectors (SNSPDs). Details of the quantum frequency conversion and detection, including conversion efficiency, filtering, and noise characterization, are described in \cite{Bock_2018, Arenskoetter2023}.

To emulate a long-distance fiber link similar to our deployed urban fiber testbed \cite{Kucera2024}, we use two 25-km fiber spools (SMF-28). One serves as the quantum channel for transmitting the frequency-converted single photons (measured transmission $24.9\,\%$), the other one (measured transmission $23\,\%$) serves as a classical channel for synchronization and detector signals. 

The quantum channel and the approximately $100$\,m of telecom fiber leading to the first QFC run through non-temperature-stabilized hallways, where environmental perturbations induce slow polarization drifts. Since polarization fluctuations would degrade the transmitted quantum correlations if left uncompensated, we employ automated polarization drift compensation (APC) based on a modified version of the scheme described in \cite{Kucera2024}. Key improvements over  \cite{Kucera2024} include the generation of polarization reference light at the ion emission wavelength of $854$\,nm and fast switching between reference polarizations using acousto-optic modulators (AOMs). To suppress residual background light during data acquisition, an additional mechanical shutter is used. During long acquisitions, the APC routine is triggered approximately every $300$\,s. If slow fiber drifts reduce the measured process fidelity below $\mathcal{F}_{\mathrm{P}}=99\,\%$, the feedback restores it above this threshold within $0.28(12)$\,s after the stabilization update starts.

Since the APC receiver operates at the telecom C-band, the polarization drifts in the $\sim100\,$m of fiber between the second QFC and the projection setup, as well as in the fibers before the APC sender stage, must be corrected separately. We achieve this using a wave-plate-based compensation scheme (APC2) with an attenuated $854\,$nm reference laser transmitted through the ion trap along the same optical path as the photons. APC2 is repeated automatically every $40\,$min, independently of the APC feedback loop. 

The classical fiber channel distributes White Rabbit synchronization \cite{Lipinski2011WhiteRabbit} signals between the time-tagging units at Alice's and Bob's stations. Between measurement runs, it also carries the APC2 detector signals from Bob to Alice. Importantly, the classical channel transmits the electronic herald signaling photon detection during measurement runs. The classical signals are transmitted simultaneously via coarse wavelength division multiplexing (CWDM).

After the second frequency conversion stage, the photons pass through a narrow-band filtering setup with a full width at half maximum of $51.17$\,MHz before entering the polarization-analysis module. This filtering, together with wavelength-selective optics along the link, strongly suppresses background detections from out-of-band light. It is also relevant for the security assumptions of the protocol, which require the trapped-ion memory to be protected against optical access via the fiber by an external adversary. Further details are provided in Appendix~\ref{App:D}.

Each SNSPD output is split electronically. One copy of each detector pulse is recorded locally by the time tagger to provide a complete detector-resolved photonic record. The other copies are combined into a setting-independent logical OR signal that is transmitted via the classical fiber channel to Alice. This electronic herald triggers the atomic readout sequence. Importantly, the herald carries no information about Bob’s measurement settings or outcomes, which are determined solely locally from the recorded detector channels.

Photon transmission from Alice to Bob, generation of the electronic herald at Bob and its transmission to Alice introduce a latency of approximately $\qty{250}{\micro\second}$. To preserve atomic coherence during this delay, the spin-echo sequence described in Sec.~\ref{sec:APE_setup} is applied. The latency also limits the repetition rate, since the ion cannot be re-initialized before the heralding and readout sequence is completed. The electronic herald rate is furthermore determined by the photon generation probability, collection efficiency, efficiencies of both frequency-conversion stages, fiber transmission, optical filtering losses, and the detection efficiency of the SNSPDs. For the optical path from the ion trap to the photonic projection setup, excluding the projection optics and SNSPD detection efficiency, we measure an overall transmission of $1.09\,\%$. At the detectors, we measure a signal-to-background ratio (SBR) of around 10.

Time-tagging of the detected photons is performed with 80\,ps resolution, which defines the fundamental time bin size, equivalent to a Larmor-phase increment of approximately $0.28^\circ$ for a Larmor period of $104.25\,$ns.

\section{QKD protocol}\label{sec:protocol}

We implement an event-ready protocol following the DIQKD setting structure of \cite{Schwonnek_2021}. Alice's measurement setting $x$ is selected locally by a quantum random number generator that chooses one of four phases of the radio-frequency $\pi/2$ pulse, corresponding to azimuthal Bloch-sphere angles of $0^\circ$, $45^\circ$, $90^\circ$, and $135^\circ$. Bob's random measurement setting $y$ is selected passively by a $50{:}50$ beam splitter that routes an incoming photon to one of two polarization-analysis modules implementing the $0^\circ$ and $90^\circ$ measurement bases. The measurement-independence assumption associated with the passive basis choice is discussed in Appendix~\ref{App:B}.

An event-ready trial that enters data analysis requires that exactly one detector clicked within a predefined narrow time window around the expected photon arrival time, and that the ion is found in the $\mathrm{D}_{5/2}$ manifold in the subsequent atomic readout (see Appendix~\ref{App:A} for details of the atomic readout and trial classification). Electronic heralds originating from clicks outside the window, for example from background detections, are therefore not counted as event-ready trials, neither are double clicks within the time window or readout results where the atom is found in $\mathrm{S}_{1/2}$ or $\mathrm{D}_{3/2}$.

Each event-ready trial is classified as either successful, if the subsequent atomic readout verifies the ion in one of the two predefined outcome states $\ket{+}$ or $\ket{-}$, or as failure otherwise. Only successful trials are used for the Bell and key analysis: successful trials with matched measurement settings ($0^\circ/0^\circ$ and $90^\circ/90^\circ$) form the raw-key data. The four combinations involving Alice's $45^\circ$ and $135^\circ$ measurement settings are used for CHSH-based parameter estimation, while the two mismatched combinations ($0^\circ/90^\circ$ and $90^\circ/0^\circ$) are not used in the present analysis. For consistency, we verified that the readout-failure probability does not depend on the measurement settings, see Appendix~\ref{App:A}.

Within the DIQKD model of \cite{Schwonnek_2021}, a CHSH value exceeding $|S|=2.362$ corresponds to a positive asymptotic secret-key fraction (SKF). We therefore extract the CHSH parameter $S$ from the atom-photon correlations and obtain the asymptotic secret-key fraction $r_\infty(|S|)$ following \cite{Schwonnek_2021}. We combine $r_\infty(|S|)$ with the number $N_{\mathrm{key}}$ of successful trials with matched measurement settings to define the asymptotic secret-key length
\[
\mathrm{SKB}=N_{\mathrm{key}} r_\infty(|S|).
\]
We denote the lower confidence bound of $\mathrm{SKB}$ by $\mathrm{SKB}_{\mathrm{low}}$ and use it to optimize the correlation-time window, as described in Sec.~\ref{sec:results}. Further details of the SKF evaluation, statistical analysis, and window optimization are provided in Appendix~\ref{App:C}.

Since the same dataset is used to both optimize the analysis and evaluate the key-related figures of merit, the reported quantities are used to characterize the performance of the present system, but not to claim device-independent security or extract a final secret key.

\section{Measurement and results}\label{sec:results}

The measurement record of a successful trial that is used for the Bell and key analysis consists of Alice's measurement setting $x$ and outcome, Bob's measurement setting $y$ and outcome, and the time tag of the photonic detection.

Alice's measurement setting is stored as an index corresponding to the four azimuthal Bloch-sphere angles $0^\circ$, $45^\circ$, $90^\circ$, and $135^\circ$. The binary outcome is obtained from state-dependent fluorescence after applying the RF $\pi/2$ rotation that selects the measurement basis. A bright fluorescence result for either of the two predefined outcome states verifies a successful trial. If neither state is verified, the trial is classified as a readout-failure trial, as defined in Sec.~\ref{sec:protocol}. 

For Bob, the measurement setting is determined by which group of two detectors, corresponding to the two polarization-analysis bases $0^\circ$ and $90^\circ$, registered the detection event. The binary outcome is then determined by which SNSPD channel in the group clicked. 

During a two-week measurement campaign, with approximately 4 days cumulative measurement uptime and 1440 million individual attempts, 28664 events satisfied the conditions for an event-ready trial. Of these, 10908 were successful and 17756 were classified as readout-failure trials. The readout-failure fraction of 61.9\% agrees well with the value expected from the branching ratios into the predefined atomic outcome states. Further details of the trial classification are provided in Appendix~\ref{App:A}.

\begin{figure}[H]
\includegraphics[width=0.48\textwidth]{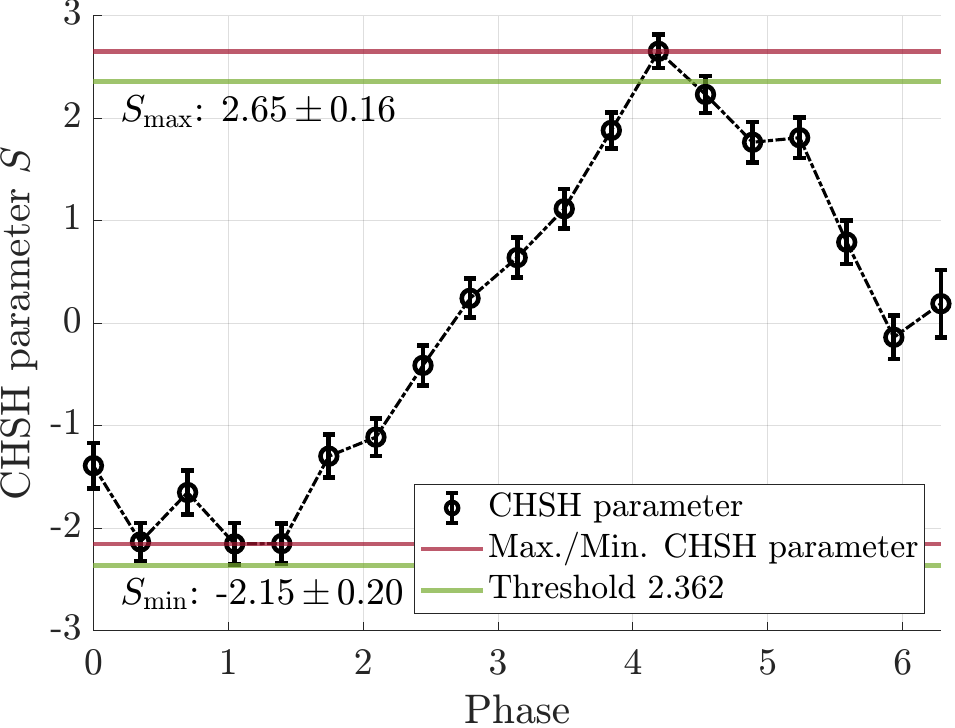}
\caption{CHSH parameter $S$ versus Larmor phase using fixed, coarse phase bins. The extrema in this representation depend on the chosen binning and are shown for visualization only. The red horizontal lines indicate the extrema, while the green horizontal lines mark the asymptotic threshold $|S|=2.362$ used in \cite{Schwonnek_2021}. The correlation-time window optimization is performed on the underlying time tags at the native $80\,\mathrm{ps}$ resolution rather than on this coarse phase-binned representation. Details of the optimization are provided in Appendix~\ref{App:C}. The quantitative results reported below are obtained from the optimized correlation-time window.}
\label{fig:raw_data_CHSH}
\end{figure}

The successful trials contain photon detections spanning the full Larmor precession period of the atomic qubit, during which the atom-photon correlation, including the CHSH parameter, oscillates with the photon detection time. Figure~\ref{fig:raw_data_CHSH} shows this approximately sinusoidal dependence using coarse Larmor-phase bins of 18$^\circ$, with the mapping from photon detection time to Larmor phase defined in Appendix~\ref{App:C}. Since the asymptotic key-fraction bound requires a large value of $|S|$, we evaluate the CHSH parameter within an optimized correlation-time window around the largest value of $|S|$ in the positive-$S$ region.

\begin{figure}[h]
		\centering
		\begin{overpic}[scale=0.65]{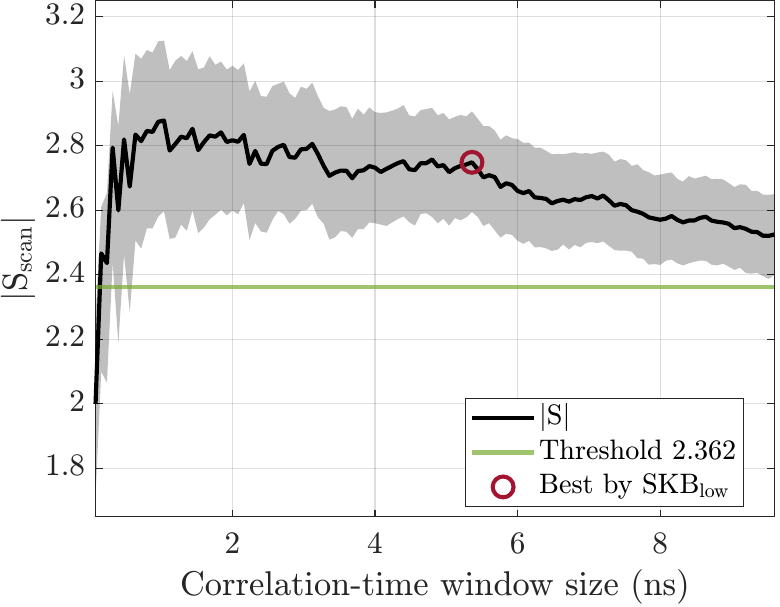}
			\put(-2,74){\large\bfseries a)}
		\end{overpic}
		\begin{overpic}[scale=0.65]{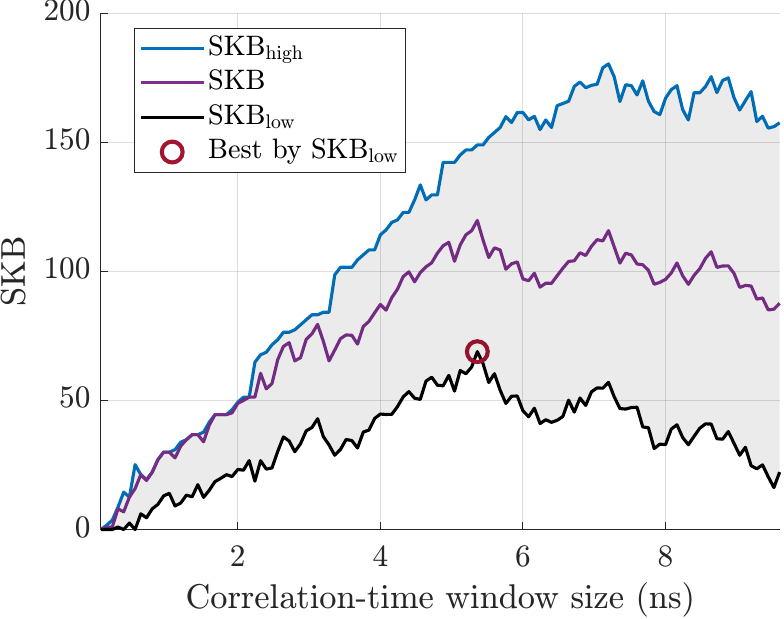}
			\put(-2,74){\large\bfseries b)}
		\end{overpic}
		\caption{Dependence of the Bell violation and asymptotic secret-key length on the correlation-time window size. For each window size, the phase offset is selected by maximizing the value of $\mathrm{SKB}$ calculated from the measured counts. (a) CHSH parameter $|S|$ for window sizes from 1 to 120 fundamental time bins corresponding to $80\,\mathrm{ps}$ to $9.6\,\mathrm{ns}$. The shaded region denotes the confidence interval defined in Appendix~\ref{App:C}, and the horizontal line marks the asymptotic threshold $|S|=2.362$ from \cite{Schwonnek_2021}. (b) Asymptotic secret-key length $\mathrm{SKB}=N_{\mathrm{key}}r_\infty(|S|)$ for the same window sizes. $\mathrm{SKB}_{\mathrm{low}}$ and $\mathrm{SKB}_{\mathrm{high}}$ denote the lower and upper confidence bounds, respectively. In both panels, the red marker denotes the window size selected by maximizing $\mathrm{SKB}_{\mathrm{low}}$.}\label{fig:window_scan}
\end{figure}

The optimized correlation-time window is found as follows: we scan window sizes from 1 to 120 fundamental time bins of 80\,ps. For each window size, we vary the phase offset to maximize the value of $\mathrm{SKB}$ calculated from the measured counts. We then select the window size that maximizes $\mathrm{SKB}_{\mathrm{low}}$ as defined in Sec.~\ref{sec:protocol}. Figure~\ref{fig:window_scan} shows the resulting dependence of $|S|$ and $\mathrm{SKB}$ on the correlation-time window size.

The optimization results in a correlation-time window of $5.36$\,ns, corresponding to 67 fundamental time bins, for which we obtain a CHSH value of $|S|=2.75^{+0.16}_{-0.15}$. This exceeds the asymptotic threshold $|S|=2.362$ of \cite{Schwonnek_2021} by approximately $2.5$ times the lower statistical uncertainty of $|S|$ (see Appendix~\ref{App:C} for evaluation of the uncertainties). For this window, we obtain a conservative estimate of $\mathrm{SKB}_{\mathrm{low}}=69$ secret-key bits under the asymptotic model.

\section{Conclusion}
This paper presents a heralded, event-ready atom-photon architecture for DIQKD and its implementation over 25\,km of fiber. By combining controlled single-photon generation with double frequency conversion and active polarization stabilization, we preserve high-fidelity atom-photon CHSH correlations across the complete link.

We employ the obtained dataset to optimize a correlation-time window in post-processing by varying its size and phase offset. At the optimized window we obtain a CHSH value of $|S|=2.75^{+0.16}_{-0.15}$, exceeding the asymptotic threshold $|S|=2.362$ of \cite{Schwonnek_2021} by approximately $2.5$ times the lower statistical uncertainty of $|S|$. Because the same dataset is used both for window optimization and for assessing the relevant figures of merit, we characterize the present system by the optimized result but do not make a device-independent key-generation claim.

Under the same asymptotic model, the optimization yields a conservative estimate of $\mathrm{SKB}_{\mathrm{low}}=69$ secret-key bits for the present conditions. Within that model, the observed Bell violation suggests that an independent run, with the analysis window fixed in advance, will reproduce a positive asymptotic key fraction.

For comparison, a local reference measurement performed without the long fiber link and the two frequency-conversion stages yields $\mathrm{SKB} \approx 5 \times 10^{4}$ over approximately 21\,h of measurement under analogous asymptotic analysis assumptions. The substantially larger value is consistent with the higher herald rate of the local setup. Fiber and frequency-conversion losses reduce the photon-detection rate in the 25-km configuration, while the approximately $250\,\mu\mathrm{s}$ feed-forward latency limits the repetition rate, because the ion cannot be re-initialized before the heralding and readout sequence is completed. 

In the future, integrated optical cavities can increase the photon collection efficiency, while segmented ion traps can enable multiplexed operations and thereby reduce the impact of the communication latency on the herald rate. In addition, excitation pulse trains \cite{Baumgart} synchronized to the Larmor period ensure that photon detections happen at a constant Larmor phase, potentially allowing all successful events to be used. Such synchronized excitation-pulse sequences have already been tested in laboratory settings.

By presenting a practical implementation that addresses the key experimental challenges, our results establish the experimental basis for extending asymmetric DIQKD architectures to deployed metropolitan fiber networks such as the Saarbr\"ucken urban fiber testbed \cite{Kucera2024}. The setup also opens up applications beyond the present DIQKD scenario. The second QFC restores the photon to the ionic wavelength, preserving compatibility with a possible $^{40}\mathrm{Ca}^{+}$-based quantum memory interface at the remote station. The same transmission architecture therefore supports direct photonic measurements as well as coupling to a further matter-based quantum-memory node, making it compatible with heterogeneous and repeater-based quantum networks.

\section*{Author contributions}
M.B. and P.B.\ set up the ion experiment.  
C.H.\ and J.M.\ designed the polarization stabilization.  
J.M., C.H., M.B. and P.B.\ performed the experiments. 
T.B. and C.B. provided the QFC technology. 
J.M. analyzed the data and wrote the manuscript with input from all authors.  
J.E.\ conceived and supervised the research.

\begin{acknowledgments}
We gratefully acknowledge support from the Federal Ministry of Research, Technology and Space (BMFTR) through projects  Q.sync (16KISQ045), QR.X (16KISQ001K) and QR.N (16KIS2180) and from the Transformationsprogramm Forschung und Wissenstransfer Saar via the Center for Quantum Technologies (QuTe). We thank Ren\'e Schwonnek for providing the numerical data used to evaluate the asymptotic secret-key fraction.
\end{acknowledgments}

\section*{Competing interests}

T.B. and C.B. are CEO and advisor, respectively, and shareholders of Optiqal Quantum Technologies GmbH, which develops quantum frequency-conversion technology. The other authors declare no competing interests.

\section*{Data availability}

The data that support the findings of this study are available from the corresponding author upon reasonable request.

\section*{Code availability}

The code used for the analysis in this study is available from the corresponding author upon reasonable request.


\appendix
\section*{Appendix}

\section{Readout-failure analysis and setting independence}
\label{App:A}

\paragraph{Trial classification.}

Each electronic herald signal triggers the subsequent atomic state readout. In the present dataset, 66575 such heralded cycles are recorded.

As first step of the atomic readout sequence, population outside the $\mathrm{D}_{5/2}$ manifold is identified by state-selective fluorescence. This check distinguishes population in the $\mathrm{S}_{1/2}$ or $\mathrm{D}_{3/2}$ manifolds from population in $\mathrm{D}_{5/2}$, but does not detect whether the ion occupies either of the two predefined outcome states in $\mathrm{D}_{5/2}$. 12904 cycles are excluded because the atomic readout identified population in $\mathrm{S}_{1/2}$ or $\mathrm{D}_{3/2}$. These cycles also include population rejected by the probabilistic balancing step (because it transfers excess population from $\mathrm{D}_{5/2}$ to $\mathrm{S}_{1/2}$). This leaves 53671 cycles.

For each remaining cycle, we search the detector-resolved time tags for exactly one photon detection within a fixed $488$\,ns time window after the earliest expected photon arrival time. This leaves $N_{\mathrm{trial}}=28664$ event-ready trials. The remaining 25007 cycles contain no detector click within this time window, i.e., the electronic herald originated from a detection outside the window, and are therefore not counted. Multiple-click events within this time window would also have been excluded, but did not occur in the present dataset.

For each event-ready trial, the two predefined atomic outcome states are subsequently evaluated using state-selective fluorescence readout. A trial is classified as successful if one of these states is verified, and as a readout-failure trial otherwise. Of the 28664 event-ready trials, $N_{\mathrm{succ}}=10908$ are successful and $N_{\mathrm{fail}}=17756$ are classified as readout-failure trials.

\paragraph{Readout-failure probability.}

Only successful trials are used for the Bell and key analysis. As a consistency check, we calculated the readout-failure probability for each measurement setting $(x,y)$ and test whether it depends on the settings.

Let $N_{\mathrm{trial}}(x,y)$ denote the number of event-ready trials for a given setting pair and $N_{\mathrm{fail}}(x,y)$ the corresponding number of readout-failure trials. We define the readout-failure probability as

\[
p_{\mathrm{fail}}(x,y)=
\frac{N_{\mathrm{fail}}(x,y)}
{N_{\mathrm{trial}}(x,y)} .
\]

The resulting values are listed in Table~\ref{tab:pfail}. Across the eight setting combinations, we observe readout-failure probabilities in the range

\[
0.6103 \le p_{\mathrm{fail}}(x,y) \le 0.6334 ,
\]

with typical binomial standard errors of approximately $0.008$.

\paragraph{Statistical test.}

To test whether these variations are compatible with statistical fluctuations, we perform a Pearson $\chi^2$ test for the null hypothesis that the readout-failure probability is independent of the measurement settings,

\[
p_{\mathrm{fail}}(x,y)=p_{\mathrm{fail}} .
\]

The test is applied to the $8\times2$ contingency table consisting of successful and readout-failure trial counts for the eight setting pairs. For the present dataset, we obtain $\chi^2=5.385$ with $7$ degrees of freedom, corresponding to a p-value of $p=0.613$, much larger than the value of 0.05 usually required for accepting the hypothesis. 

Although the values of $p_{\mathrm{fail}}(x,y)$ differ slightly between setting combinations, the Pearson $\chi^2$ test shows that these variations are consistent with the statistical fluctuations expected from the finite sample size. We therefore find no statistically significant dependence of the readout-failure probability on the measurement settings. This consistency check, however, does not establish a fair-sampling condition or close the detection loophole.

\begin{table}[t]
\centering
\caption{Readout-failure probability $p_{\mathrm{fail}}(x,y)$ for all measurement-setting combinations. The mapping from the setting labels $x\in\{0,1,2,3\}$ and $y\in\{0,1\}$ to physical measurement bases is defined in Sec.~\ref{sec:protocol} and Appendix~\ref{App:B}. Uncertainties correspond to $1\sigma$ binomial standard errors.}
\label{tab:pfail}
\begin{tabular}{c|cc}
\hline
Alice setting $x$ & $y=0$ (a/b, $90^\circ$) & $y=1$ (c/d, $0^\circ$) \\
\hline
0 ($0^\circ$)   & $0.6202 \pm 0.0081$ & $0.6181 \pm 0.0081$ \\
1 ($45^\circ$)  & $0.6103 \pm 0.0081$ & $0.6228 \pm 0.0081$ \\
2 ($90^\circ$)  & $0.6193 \pm 0.0081$ & $0.6334 \pm 0.0080$ \\
3 ($135^\circ$) & $0.6117 \pm 0.0081$ & $0.6200 \pm 0.0081$ \\
\hline
\end{tabular}
\end{table}

\section{Setting generation and measurement-independence assumption}\label{App:B}

We assume that Alice's locally generated setting and Bob's passive setting choice are statistically independent of the quantum state distributed to the measurement devices. This corresponds to the standard measurement-independence assumption used in device-independent and Bell-test-based security analyses. The beam splitter and all associated routing optics are located inside Bob's secure laboratory and are not accessible to an external adversary.

\section{Correlation-time window and statistical analysis}\label{App:C}

Appendix~\ref{App:A} describes how event-ready trials are selected. Only successful trials enter the Bell and key analysis. Here we describe how the phase-dependent correlations are evaluated and how the correlation-time window is selected for future fixed-window operation. The correlation-time window is applied only in post-processing and does not affect the classification of event-ready trials. Because the optimization is performed on the same dataset used to report the resulting quantities, these optimized-window results are used only to characterize the present system and do not constitute a device-independent security claim.

\paragraph{Larmor-phase mapping and window parameterization.}

Photon detection timestamps are discretized at the time-tagger resolution of $80\,\mathrm{ps}$. Each photon-detection time $t$ relative to the excitation trigger is mapped to a Larmor phase $\phi\in[0,2\pi)$ according to
\[
\phi = \mathrm{mod}\!\left(2\pi f_L t,2\pi\right),
\]
using the independently calibrated Larmor frequency $f_L = 9.59(1)\,\mathrm{MHz}$.

For each window size $w$, expressed in units of the fundamental $80$\,ps time bins, we scan all possible phase offsets of the correlation-time window over one Larmor period with the same $80$\,ps resolution. Each candidate correlation-time window is therefore defined by its size $w$ and its phase offset, both in units of 80-ps bins.

\paragraph{CHSH estimator.}

For each candidate correlation-time window, we calculate from the data within the window the correlation coefficient
\[
E(x,y)=
\frac{n_{++}+n_{--}-n_{+-}-n_{-+}}
{n_{++}+n_{--}+n_{+-}+n_{-+}},
\]
where $n_{ab}$ denotes the number of successful trials with Alice outcome $a\in\{+,-\}$ and Bob outcome $b\in\{+,-\}$ for the measurement-setting pair $(x,y)$. The CHSH parameter is calculated as
\[
S =
E(45^\circ,90^\circ)
+
E(135^\circ,90^\circ)
+
E(45^\circ,0^\circ)
-
E(135^\circ,0^\circ).
\]
For the reported values, the correlation coefficients $E(x,y)$ in this expression are replaced by the regularized coefficients $E_\beta(x,y)$ defined below.

To avoid unstable values when the number of trials within a correlation-time window is small, we apply a weak regularization to the correlation coefficients. For an outcome-count vector $(n_{++},n_{+-},n_{-+},n_{--})$, the regularized correlation coefficient is evaluated as
\[
E_{\beta}(x,y)=
\frac{n_{++}+n_{--}-n_{+-}-n_{-+}}
{n_{++}+n_{--}+n_{+-}+n_{-+}+2\beta}.
\]
This is equivalent to adding a pseudocount of $\beta/2$ to each of the four outcome counts, which leaves the numerator unchanged and adds $2\beta$ to the denominator. For the reported CHSH values, we use $E_\beta(x,y)$ with $\beta=1$ in the CHSH expression above. This regularization suppresses unstable correlation estimates in the low-count regime, including values close to $\pm1$ that can arise from very sparse outcome counts, while its effect becomes negligible for larger samples.

\paragraph{Robustness check for the correlation regularization.}

To assess the sensitivity of the analysis to the regularization used in the low-count regime, we repeat the complete window optimization for $\beta\in\{0,1,2\}$. In all three cases, the same correlation-time window is selected. The corresponding values of $|S|$ vary only weakly compared with the quoted statistical uncertainty, indicating that the reported result is not driven by the choice of $\beta$ within this range.

\paragraph{Asymptotic secret-key fraction and secret-key length.}

We evaluate the asymptotic secret-key fraction $r_\infty(|S|)$ using the numerical data provided for the respective protocol of Ref.~\cite{Schwonnek_2021}. In the depolarizing-noise model used there, the QBER entering the key evaluation is related to the CHSH value by
\[
Q(S)=\frac{1}{2}\left(1-\frac{|S|}{2\sqrt{2}}\right).
\]
Thus, for each measured value of $|S|$, we first determine the corresponding QBER $Q(S)$ and then obtain the associated asymptotic secret-key fraction $r_\infty(|S|)$ from the numerical data of Ref.~\cite{Schwonnek_2021}. 

When evaluating $r_\infty(|S|)$, values of $|S|$ outside the tabulated range are restricted to the nearest boundary value, and negative values of $r_\infty$ are set to zero. In particular, values $|S|>2\sqrt{2}$ are set to the physical CHSH bound $2\sqrt{2}$.

The corresponding asymptotic secret-key length is defined as
\[
\mathrm{SKB}=N_{\mathrm{key}}r_\infty(|S|),
\]
where $N_{\mathrm{key}}$ is the number of successful matched-basis trials in the $0^\circ/0^\circ$ and $90^\circ/90^\circ$ measurement settings that fall within the same correlation-time window.

In the present analysis, we use $\mathrm{SKB}$ as a model-dependent performance metric and to compare candidate correlation-time windows. It does not represent an extracted secret key. For the selected correlation-time window, $N_{\mathrm{key}}=154$ successful matched-basis trials enter the key-length calculation, while 317 successful trials contribute to the CHSH estimation. The remaining 158 successful trials within the window belong to the two mismatched setting combinations that are not used in the present analysis.

\paragraph{Poisson bootstrap.}

In the low-count regime, Gaussian error propagation and symmetric $\sqrt{N}$ uncertainties are unreliable for nonlinear quantities such as the CHSH parameter, which is calculated from ratios of count variables. We therefore estimate statistical uncertainties using a Poisson bootstrap procedure.

In each bootstrap realization, every successful trial contributing to the CHSH analysis is assigned a random, independent Poisson-distributed weight $w_i$ with mean one,
\[
w_i \sim \operatorname{Poisson}(1).
\]
A weight of zero removes a trial from that bootstrap realization, a weight of one retains it once, and larger weights include it correspondingly multiple times.

The weighted detector-resolved trial sample is then used to recalculate the correlation coefficients and the CHSH parameter $S$.

For the corresponding $\mathrm{SKB}$ distribution, the observed number of successful matched-basis trials $N_{\mathrm{key}}$ within the selected correlation-time window is kept fixed, while the secret-key fraction $r_\infty(|S|)$ is recalculated for each bootstrap realization. The bootstrap therefore propagates the statistical uncertainty of the CHSH estimate into $r_\infty$ and $\mathrm{SKB}$, while conditioning on the observed value of $N_{\mathrm{key}}$. The resulting interval therefore quantifies the statistical uncertainty of the inferred secret-key fraction and the corresponding SKB for the observed value of $N_{\mathrm{key}}$.

\paragraph{Bootstrap intervals and window-size optimization.}

We report intervals obtained directly from percentiles of the bootstrap distribution rather than assuming a symmetric distribution. For a central confidence level $1-\alpha$, with $\alpha=0.3173$ for the reported $68.27\%$ interval, we use the $\alpha/2$ and $1-\alpha/2$ percentiles in accordance with standard bootstrap percentile intervals \cite{EfronTibshirani1993Bootstrap,DiCiccioEfron1996BootstrapCI}.

For each correlation-time window size, we first scan all phase offsets and select the offset that maximizes the value of $\mathrm{SKB}$ calculated from the measured counts. For this selected offset, we then calculate the bootstrap distributions of $S$ and $\mathrm{SKB}$ and define $\mathrm{SKB}_{\mathrm{low}}$ as the lower bound of the central $68.27\%$ bootstrap interval. Finally, among all tested window sizes, we select the window size with the largest value of $\mathrm{SKB}_{\mathrm{low}}$. This favors window sizes whose performance is robust to statistical fluctuations rather than those selected solely by a large value of $\mathrm{SKB}$ calculated from the measured counts.

The bootstrap intervals quantify the statistical uncertainty for a fixed window whose size and phase offset have been selected from the measured data. They do not account for the selection bias introduced by choosing this window from the large number of scanned candidates. The optimized result is therefore not interpreted as a device-independent security result.

\section{Optical isolation of the trapped-ion memory}\label{App:D}

\paragraph{Scope of the assumption.}

Beyond the assumptions considered in the present analysis, we assume standard laboratory isolation, namely that an external adversary has no practical optical access to the trapped-ion apparatus through the fiber link. Measures that support this assumption are specified below. No quantitative security bound on this optical isolation is derived here.

\paragraph{Wavelength separation and spectral isolation.}

The quantum channel is designed to transmit only photons originating from the ion's $854\,\mathrm{nm}$ emission. These photons are converted to the telecom C-band for transmission and converted back to $854\,\mathrm{nm}$ prior to polarization analysis. The conversion stages, together with wavelength-selective optics along the link and the narrowband filter after the second converter with a FWHM of $51.17\,\mathrm{MHz}$, define a narrow spectral acceptance window around the $854\,\mathrm{nm}$ signal and strongly suppress out-of-band light.

An adversary could inject bright light into the fiber link, in particular at telecom wavelengths such as $1550\,\mathrm{nm}$. Telecom-wavelength light is not efficiently converted to a wavelength resonant with the ion, and residual light propagating through the conversion chain is further suppressed by the wavelength-selective optics and narrowband filtering around $854\,\mathrm{nm}$.

The wavelengths used for coherent control and state readout of the $^{40}\mathrm{Ca}^{+}$ ion are generated and applied locally at Alice. These include $393\,\mathrm{nm}$ for excitation, $397\,\mathrm{nm}$ for fluorescence detection, $866\,\mathrm{nm}$ for repumping, and $729\,\mathrm{nm}$ for coherent state transfer. These wavelengths lie outside the acceptance bands of the frequency-conversion and filtering stages and are therefore strongly attenuated along the implemented quantum-channel path.

\paragraph{Heralding, feed-forward, and absence of an optical back-channel.}

The electronic herald signal is generated from the SNSPD detection signals at Bob and transmitted to Alice through the classical fiber channel. This signal initiates the subsequent atomic readout sequence. Event-ready trials are selected separately from the photon detection record as described in Appendix~\ref{App:A}.

No optical signal is sent from Bob to Alice through the quantum channel as part of the heralding or feed-forward procedure. The heralding information is carried by the separate classical fiber channel and converted to an electronic control signal at Alice before entering the atomic control sequence. The heralding path therefore does not provide a direct optical path to the trapped-ion system.

\paragraph{Memory state and exposure time.}

During an event-ready trial, the atomic qubit is stored in the metastable $\mathrm{D}_{5/2}$ manifold until the subsequent atomic readout. After the readout, the ion is re-initialized before the next measurement sequence. Influencing or probing the atomic memory through the fiber link would require light to reach the ion with appropriate wavelength and sufficient power. In the present implementation, such access is strongly suppressed by the spectral selectivity and directivity of the frequency-conversion stages and filtering optics and by the absence of a direct optical path from the heralding channel to the trapped-ion system.

\bibliography{Bibliography.bib}

\end{document}